# Fastest Distributed Consensus Problem on Branches of an Arbitrary Connected Sensor Network

Saber Jafarizadeh, *Student Member, IEEE,* and Abbas Jamalipour, *Fellow, IEEE*

***Abstract*—This paper studies the fastest distributed consensus averaging problem on branches of an arbitrary connected sensor network. In the previous works full knowledge about the sensor network's connectivity topology was required for determining the optimal weights and convergence rate of distributed consensus averaging algorithm over the network. Here in this work for the first time, the optimal weights are determined analytically for the edges of certain types of branches, independent of the rest of network. The solution procedure consists of stratification of associated connectivity graph of the branches and Semidefinite Programming (SDP), particularly solving the slackness conditions, where the optimal weights are obtained by inductive comparing of the characteristic polynomials initiated by slackness conditions. Several examples and numerical results are provided to confirm the optimality of the obtained weights.**

## I. INTRODUCTION

DISTRIBUTED consensus has appeared as one of the important and primary problems in the context of distributed computation (see, for example, [1] for early work). Some of its applications include distributed agreement, synchronization problems, [2] and load balancing in processor networks [3, 4].

A problem that has received renewed interest recently is distributed consensus averaging algorithms in sensor networks. The main purpose of distributed consensus averaging algorithm on a sensor network is to compute the average of the initial node values, via a distributed algorithm, in which the nodes only communicate with their immediate neighbors. One of main research directions in this issue is the computation of the optimal weights that yield the fastest convergence rate to the asymptotic solution [5, 6] known as Fastest Distributed Consensus (FDC) averaging algorithm.

Moreover algorithms for distributed consensus find applications in, e.g., multi-agent distributed coordination and flocking [7, 8, 9], distributed data fusion in sensor networks [10, 11, 6], fastest mixing Markov chain problem [12], gossip algorithms [13, 14], and distributed estimation and detection for decentralized sensor networks [15, 16, 17, 18, 19].

In previous works determining optimal weights and convergence rate of FDC averaging algorithm over a sensor network demanded full knowledge about network's topology. But here in this work we have determined the optimal weights for edges of five different branches connected to an arbitrary network without requiring full knowledge about network's topology. The branches considered in this paper are path, lollipop, semi-complete, ladder and palm branches. Stratification [20, 21] and semidefinite programming are the key methods used for evaluating optimal weights. We have proved that the obtained weights are optimal and independent of rest of network. Several examples and simulation results are provided to confirm the validity of the obtained results. By numerical simulations we have compared the branches in terms of asymptotic and per step convergence rates. Also it is shown that the obtained results hold true in serial combination of these branches too.

The organization of the paper is as follows. In section II we briefly review prior works on the agreement and consensus algorithms. Section III is an overview of the materials used in the development of the paper, including relevant concepts from distributed consensus averaging algorithm, graph symmetry, stratification and semidefinite programming. Section IV contains the main results of the paper where four different branches namely path, lollipop, semi-complete, ladder and palm branches are introduced together with the obtained optimal weights. In section V some examples are presented. Section VI is devoted to the proof of main results of paper. Section VII presents simulations and section VII concludes the paper.

## II. RELATED WORKS

Tsitsiklis [1] gave a systematic study of agreement algorithms of the general type in an asynchronous distributed environment. Their recent work [9] summarizes the key results and establishes some new extensions. In [8], a continuous time state update model was adopted for consensus and the results were extended to situations involving switching sensor network topology and delayed communication. In [12] it has been shown that the convergence rate of FDC averaging algorithm is determined by the second largest eigenvalue modulus (SLEM) of the weight matrix furthermore FDC averaging problem is formulated as a convex optimization problem, in particular a semidefinite program.

A well-studied method for choosing weights is the nearest neighbor rule. This method of choosing weights was analyzed in detail in [7]. In [5], the problem of designing the optimal weights was addressed for a fixed sensor network topology.

S. Jafarizadeh and A. Jamalipour are with the School of Electrical and Information Engineering, University of Sydney, Sydney NSW 2006, Australia (e-mail: saber.jafarizadeh@sydney.edu.au & abbas.jamalipour@sydney.edu.au).

The authors deal with the FDC averaging problem by numerical convex optimization methods but no closed-form solution for finding the optimal weights were proposed. In [6] the authors introduce the Metropolis weights over the time-varying communication graphs. They show that Metropolis weights preserve the average of node values and converge to the average of node values, provided the infinitely occurring communication graphs are jointly connected. In [20] the authors show how to exploit symmetries of a graph to efficiently compute the optimal weights of the fastest distributed consensus averaging algorithm.

Many works dealt with reaching a quantized consensus (see [22, 23] and references therein). In [24], Carli et al. consider the problem of reaching a consensus using quantized channels. They propose to round each node value at each step to the nearest integer. The proposed quantization method by [24] conserves the average of the states but the nodes do not necessarily reach a consensus and they converge to different values.

In [25], Aysal et al. propose a different method for quantization called "probabilistic quantization" scheme. Using their method nodes converge to a consensus but the average of node values is not conserved.

## III. Preliminaries

This section introduces the notation used in the paper and reviews relevant concepts from distributed consensus averaging algorithm, stratification, graph symmetry and semidefinite programming.

### A. Distributed Consensus Averaging Algorithm

We consider a network $\mathcal{N}$ with the associated graph $\mathcal{G} = (\mathcal{V}, \mathcal{E})$ consisting of a set of nodes $\mathcal{V}$ and a set of edges $\mathcal{E}$ where each edge $\{i,j\} \in \mathcal{E}$ is an unordered pair of distinct nodes.

The main purpose of distributed consensus averaging is to compute the average of the initial node values $\bar{x} = (\mathbf{1}\mathbf{1}^T/n)x(0)$, through the distributed linear iterations $x(t+1) = Wx(t)$. $x(0)$ is the vector of initial node values on the network. $W$ is the weight matrix with the same sparsity pattern as the adjacency matrix of the network's associated graph and $t = 0,1,2,\ldots$ is the discrete time index (Here $\mathbf{1}$ denotes the column vector with all coefficients one).

In [5] it has been shown that the necessary and sufficient conditions for the convergence of linear iteration mentioned above is that one is a simple eigenvalue of $W$ associated with the eigenvector $\mathbf{1}$, and all other eigenvalues are strictly less than one in magnitude. Moreover in [5] FDC averaging problem has been formulated as the following minimization problem

$$\begin{aligned} &\min_{W} \quad \max(\lambda_2, -\lambda_n) \\ &s.t. \quad W = W^T, W\mathbf{1} = \mathbf{1}, \forall \{i,j\} \notin \mathcal{E}: W_{ij} = 0 \end{aligned}$$

where $1 = \lambda_1 \geq \lambda_2 \geq \cdots \geq \lambda_n \geq -1$ are eigenvalues of $W$ arranged in decreasing order and $\max(\lambda_2, -\lambda_n)$ is the *Second Largest Eigenvalue Modulus* (*SLEM*) of $W$, and the main problem can be formulated in the semidefinite programming form as [5]:

$$\begin{aligned} &\min_{W} \quad s \\ &s.t. \quad -sI \preccurlyeq W - \mathbf{1}\mathbf{1}^T/n \preccurlyeq sI, W = W^T \\ &\qquad W\mathbf{1} = \mathbf{1}, \forall \{i,j\} \notin \mathcal{E}: W_{ij} = 0. \end{aligned} \tag{1}$$

We refer to problem (1) as the Fastest Distributed Consensus (FDC) averaging problem.

### B. Stratification & Symmetry of Graphs

An automorphism of a graph $\mathcal{G} = (\mathcal{V}, \mathcal{E})$ is a permutation $\sigma$ of $\mathcal{V}$ such that $\{i,j\} \in \mathcal{E}$ if and only if $\{\sigma(i), \sigma(j)\} \in \mathcal{E}$, the set of all such permutations, with composition as the group operation, is called the automorphism group of the graph and denoted by $Aut(\mathcal{G})$. For a vertex $i \in \mathcal{V}$, the set of all images $\sigma(i)$, as $\sigma$ varies through a subgroup $G \subseteq Aut(\mathcal{G})$, is called the orbit of $i$ under the action of $G$. The vertex set $\mathcal{V}$ can be written as disjoint union of distinct orbits.

Stratifying a graph into its orbits is called stratification [20, 21] (in [20] this method is entitled block-diagonalization) and each orbit is called a stratum. In [20], it has been shown that the weights on the edges within an orbit are the same. For the sake of clarity, consider the wheel graph; $W_7$ (see Fig. 1).

The automorphism group $Aut(G)$ is isomorphic to the 6-element cyclic group $C_6$, and corresponds to flips and 60-degree rotations of the graph. The orbits of $Aut(G)$ acting on the vertices are

$$\{1,2,3,4,5,6\}, \{0\}$$

and there are two orbits of edges

$$\big\{\{0,1\},\{0,2\},\{0,3\},\{0,4\},\{0,5\},\{0,6\}\big\},$$
$$\big\{\{1,2\},\{2,3\},\{3,4\},\{4,5\},\{5,6\},\{6,1\}\big\}.$$

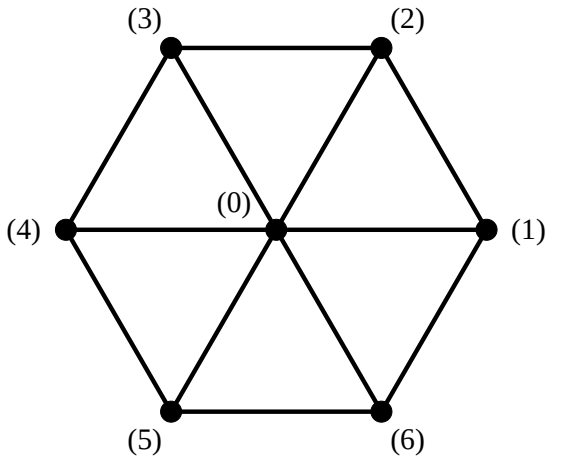

Fig. 1. a wheel graph ($W_7$).

### C. Semidefinite Programming (SDP)

SDP is a particular type of convex optimization problem [26]. An SDP problem requires minimizing a linear function subject to a linear matrix inequality constraint [27]:

$$\begin{aligned} &\min \quad \rho = c^T x, \\ &s.t. \quad F(x) \geq 0 \end{aligned}$$

where $c$ is a given vector, $x^T = (x_1, \ldots, x_n)$, and $F(x) = F_0 + \sum_i x_i F_i$, for some fixed hermitian matrices $F_i$. The inequality sign in $F(x) \geq 0$ means that $F(x)$ is positive semi-definite.

This problem is called the primal problem. Vectors $x$ whose components are the variables of the problem and satisfy the constraint $F(x) \geq 0$ are called primal feasible points, and if they satisfy $F(x) > 0$, they are called strictly feasible points. The minimal objective value $c^T x$ is by convention denoted by $\rho^*$ and is called the primal optimal value.

Due to the convexity of the set of feasible points, SDP has a nice duality structure, with the associated dual program being:

$$\begin{aligned} \max \quad & -Tr[F_0 Z] \\ s.t. \quad & Z \geq 0, Tr[F_i Z] = c_i \end{aligned}$$

Here the variable is the real symmetric (or Hermitian) positive matrix $Z$, and the data $c$, $F_i$ are the same as in the primal problem. Correspondingly, matrix $Z$ satisfying the constraints is called dual feasible (or strictly dual feasible if $Z > 0$). The maximal objective value of $-Tr[F_0 Z]$, i.e. the dual optimal value is denoted by $d^*$.

The objective value of a primal (dual) feasible point is an upper (lower) bound on $\rho^*(d^*)$. The main reason why one is interested in the dual problem is that one can prove that $d^* \leq \rho^*$, and under relatively mild assumptions, we can have $\rho^* = d^*$. If the equality holds, one can prove the following optimality condition on $x$.

A primal feasible $x$ and a dual feasible $Z$ are optimal, which is denoted by $\hat{x}$ and $\hat{Z}$, if and only if

$$F(\hat{x})\hat{Z} = \hat{Z}F(\hat{x}) = 0. \tag{2}$$

which is called the complementary slackness condition.

In one way or another, numerical methods for solving SDP problems always exploit the inequality $d \leq d^* \leq \rho^* \leq \rho$, where $d$ and $\rho$ are the objective values for any dual feasible point and primal feasible point, respectively. The difference $\rho^* - d^* = c^T x + Tr[F_0 Z] = Tr[F(x)Z] \geq 0$ is called the duality gap. If the equality $d^* = \rho^*$ holds, i.e. the optimal duality gap is zero, then we say that strong duality holds.

## IV. Optimal Weights

Here we have introduced five different branches namely Path branch, Lollipop branch, Semi-complete branch, Ladder branch and palm branch along with their corresponding evaluated optimal weights. Proofs and more detailed discussion are deferred to Section VI.

### A. Path Branch

The simplest form of branch is Path branch where a path graph $P_n$ consisting of $n$ nodes is connected to an arbitrary graph by a bridge as shown in Fig. 2.

Using the same procedure as done in section VI for semi-complete branch we can state that the optimal weights for the edges of a path branch of an unknown network, equals $1/2$, independent of the rest of network, except for the last edge (weighted by $w_n$ in Fig. 2) which connects path branch to the rest of the network.

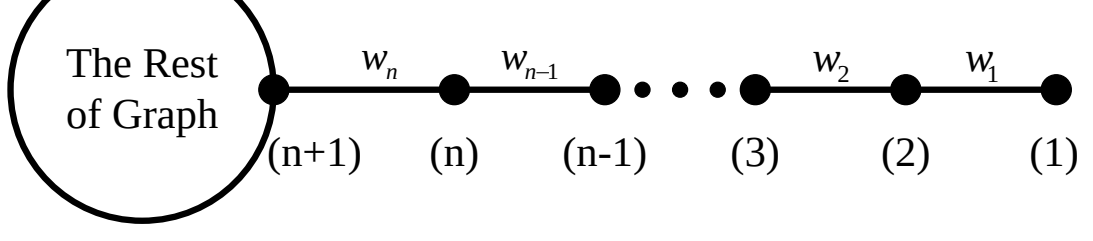

Fig. 2. An arbitrary graph with a path branch $P_n$.

### B. Lollipop Branch

The $(m,n)$-lollipop graph $L_{m,n}$ is the graph obtained by joining a complete graph $K_m$ to a path graph $P_n$. We define the lollipop branch $L_{m,n}$ as a lollipop graph which is connected to an arbitrary graph by sharing the end node of path part with the arbitrary graph as shown in Fig. 3 for $n = 3$, $m = 5$.

In section VI-A we have proved that in a Lollipop branch of an arbitrary network, the optimal weights for the edges on the complete graph $K_m$ and path graph $P_n$ equal $1/m$ and $1/2$, respectively, independent of the rest of network, except for the last edge which connects Lollipop branch to the rest of the network (weighted by $w_3$ in Fig. 3).

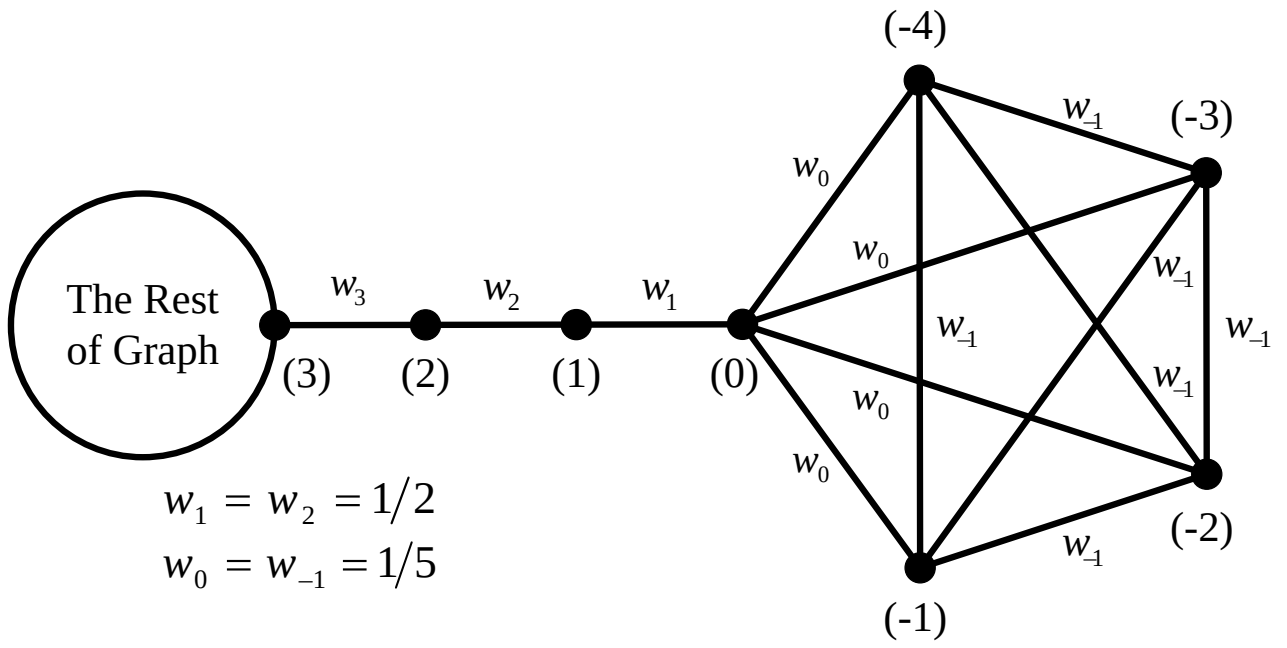

Fig. 3. An arbitrary graph with a Lollipop branch $L_{m,n}$ for $n = 3$, $m = 5$.

### C. Semi-Complete Branch

The $(m, n_1, n_2)$ Semi-Complete branch is a path branch with a semi-complete graph inside. A semi-complete graph is a complete graph without the edge between two nodes connected to path links. The whole branch consists of two path graphs $P_{n_1}$ and $P_{n_2}$ with $n_1$ and $n_2$ nodes, respectively, where each one of path graphs are connected to semi-complete graph $K_m$ by means of a bridge, as shown in Fig. 4 for $n_1 = 2, n_2 = 4, m = 5$.

In section VI-B it has been proved that in a semi-complete branch of an arbitrary network, the optimal weights on the edges connecting inner nodes together and the edges connecting outer nodes to inner nodes in the semi-complete graph equal $\frac{m-3}{(m-1)(m-2)}$ and $1/(m-1)$, respectively. Outer nodes of semi-complete graph are the two nodes connected to path graphs by a bridge and the remaining $(m-2)$ nodes of semi-complete graph are the inner nodes of semi-complete graph. The optimal weights on the edges of path graphs $P_{n_1}$ and $P_{n_2}$ equal $1/2$ independent of the rest of network, except the last edge which connects the semi-complete branch to the rest of the network (weighted by $w_9$ in Fig. 4).

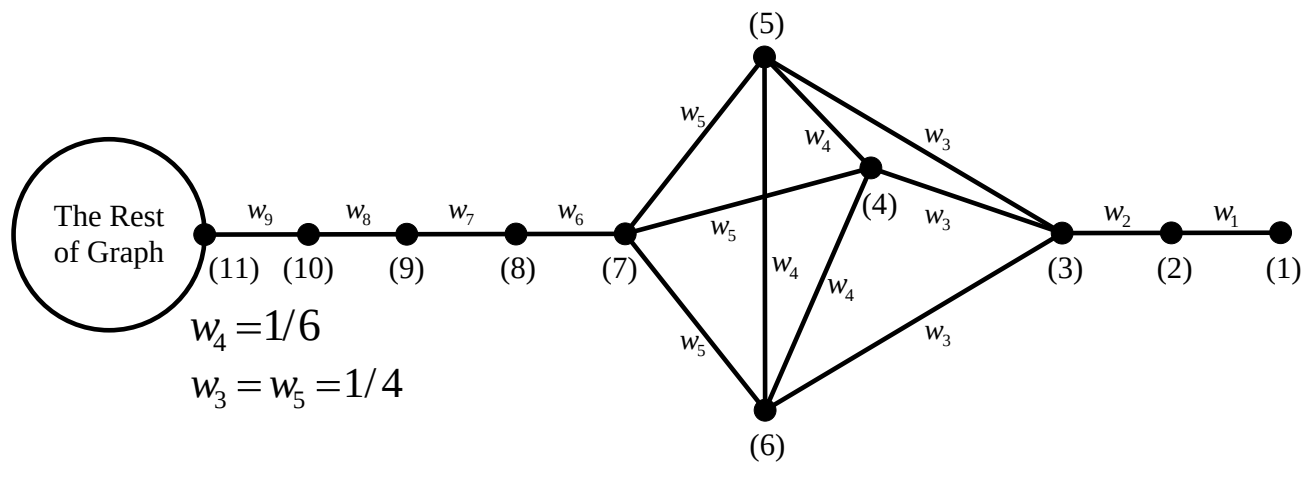

Fig. 4. An arbitrary graph with a semi-complete branch for $n_1 = 2$, $n_2 = 4$, $m = 5$.

### D. Ladder Branch

The $(m, n_1, n_2)$ Ladder branch consists of two path branches $P_{n_1+1}$ and $P_{n_2+1}$ connected by four bridges to a complete ladder graph. A complete ladder graph $CL_m$ is a ladder graph with $2(m+1)$ nodes including a complete graph $K_4$ between every 4 neighboring nodes, or in other words a complete ladder graph $CL_m$ consists of $m$ complete graphs $K_4$ where everyone of these complete graphs are sharing two nodes with the neighboring ones. A $(m, n_1, n_2)$ Ladder branch is depicted in Fig. 5.

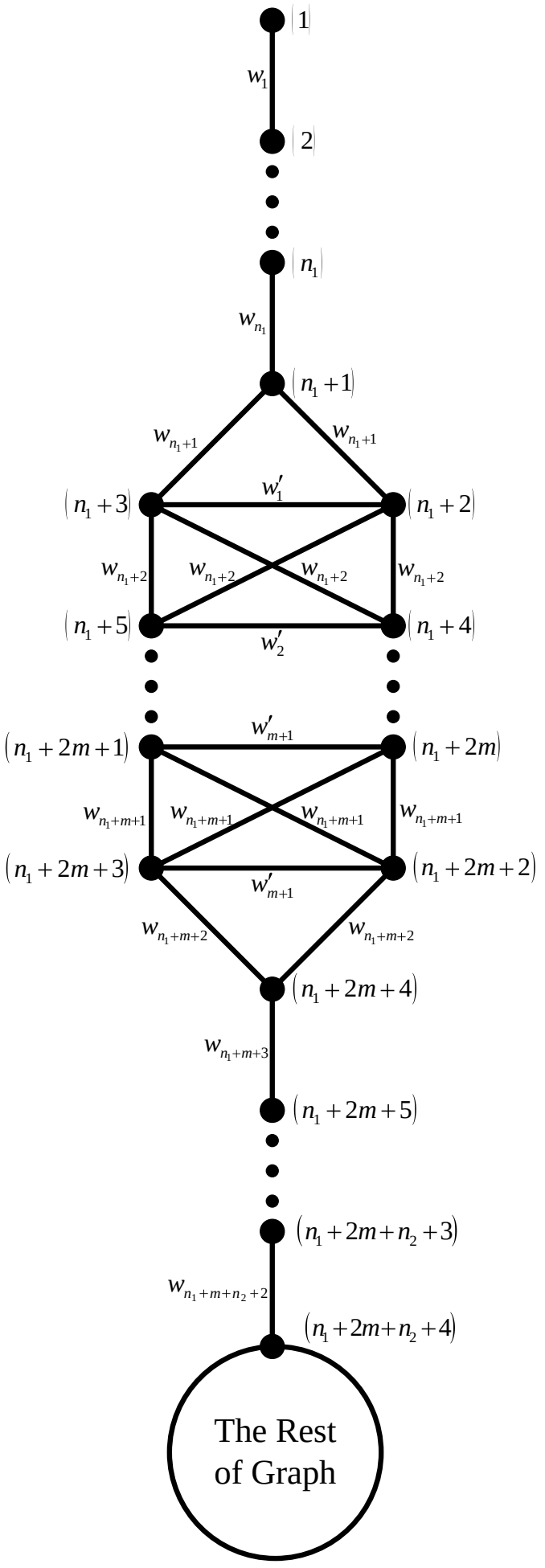

Fig. 5. An arbitrary graph with a $(m, n_1, n_2)$ Ladder branch.

In section VI-C it has been proved that in a Ladder branch of an arbitrary network, the optimal weights of the edges connecting two nodes from one level of ladder to the nodes on the neighboring levels equal $1/4$ where the nodes on one level of complete ladder graph are the nodes on the same vertical position in Fig. 5. The complete ladder graph $CL_m$ includes $(m+1)$ levels of nodes. The optimal weights of four bridges connecting complete ladder graph $CL_m$ to two path graphs $P_{n_1+1}$ and $P_{n_2+1}$ equal $1/3$. The optimal weights of the edges of path graphs $P_{n_1+1}$ and $P_{n_2+1}$ equal $1/2$, independent of the rest of network, except the last edge which connects the semi-complete branch to the rest of the network (weighted by $w_{n_1+2m+n_2+4}$ in Fig. 5). Also in section VI-C it has been shown that the *SLEM* of network is independent of the weights of edges connecting two nodes of complete ladder graph $CL_m$ on the same level to each other (weighted by $w'_i$ for $i = 1, \dots, m+1$ in Fig. 5) as long as these weights satisfy (20).

## E. *Palm Branch*

The palm branch of order $(m, k)$ is obtained by joining a path graph of length $m$ to a symmetric star graph with $k$ branches of length 1 as shown in Fig. 6 for $m = 3$, $k = 4$.

In section VI-D we have proved that in a palm branch of an arbitrary network, the optimal weights for the edges on path graph and star graph equal $1/2$ and $1/(k+1)$ , respectively, independent of the rest of network, except for the last edge which connects Lollipop branch to the rest of the network (weighted by $w_3$ in Fig. 6).

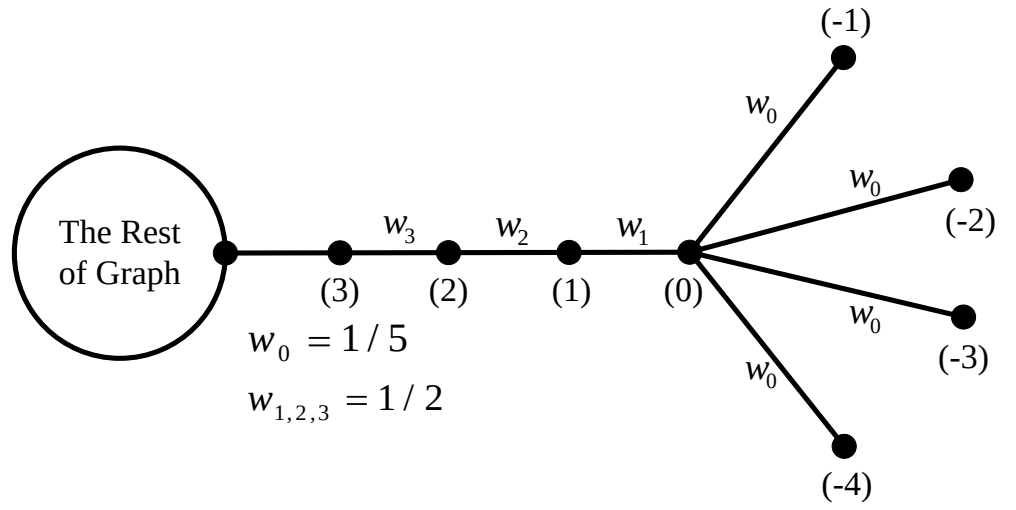

Fig. 6. An arbitrary graph with a palm branch for $m = 3$, $k = 4$.

# V. EXAMPLES

In this section several examples for five types of branches introduced in previous section are presented.

## A. *Path Network*

The simplest example is a sensor network with path topology. In a path network, the paths from every middle node to both ends of network can be considered as a path formed branch with the optimal weights equal $1/2$ which agrees with those of [28].

## B. *Star, Complete Cored Star & Two Fused Star Networks*

As another example, in [29, 30] three types of networks have been studied, namely, Complete Cored Star, Star and Two Fused Star networks where in these networks path formed branches are connected to a complete graph and a central node respectively. In [29, 30] it has been proved that the optimal weights for the edges on path formed branches are $1/2$ except for the last edge which connects the branch to the rest of network which confirms the results obtained in section VI.

## C. *Extended Barbell Network*

As an example for the Lollipop branch we consider the extended Barbell topology which is a network obtained by connecting two networks with complete graph topology $K_{m_1}$, $K_{m_2}$ by a path bridge $P_{n-1}$ as shown in Fig. 7 for $m_1 = 5, m_2 = 4$ and $n = 3$.

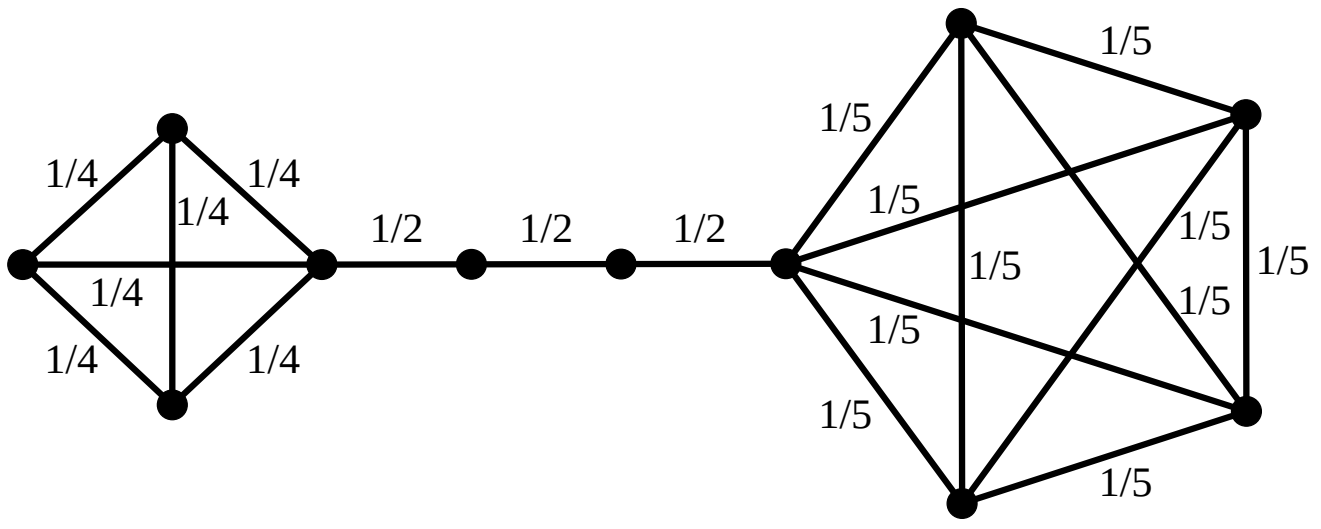

Fig. 7. Barbell graph for $m_1 = 5, m_2 = 4$ and $n = 3$.

Dividing an extended Barbell network from every node on path bridge splits the network into two Lollipop branches and the results obtained in section VI-B imply that the optimal weights equal $1/2$ for the edges of path bridge $P_{n-1}$ and $1/m_1, 1/m_2$ for the edges of each one of complete graphs $K_{m_1}$ and $K_{m_2}$, respectively.

According to subsection III-B, the stratification of Barbell network reduces it to a path graph $P_{n+3}$ with $n+3$ nodes and $m_1+m_2-4$ disconnected single nodes where $n$ is the number of edges on the path bridge. Using the results obtained in [28] and section VI-B, one can deduce that the weight matrix of extended Barbell network has $m_1+m_2-4$ zero eigenvalues and $n+3$ non zero eigenvalues, which are $\cos(k\theta)$ for $k=1,\dots,n+2$ with $\theta=\pi/(n+3)$ and one eigenvalue equal to 1. Thus the *SLEM* of Barbell network equals $\cos(\pi/n+3)$.

### D. Semi-Complete Network with Two Path branches

As a simple example for the semi-complete branch introduced in section IV-B we can consider a semi-complete graph $K_m$ with two path branches $P_{n_1}$ and $P_{n_2}$ as depicted in Fig. 8 for $n_1=2, n_2=3$ and $m=5$.

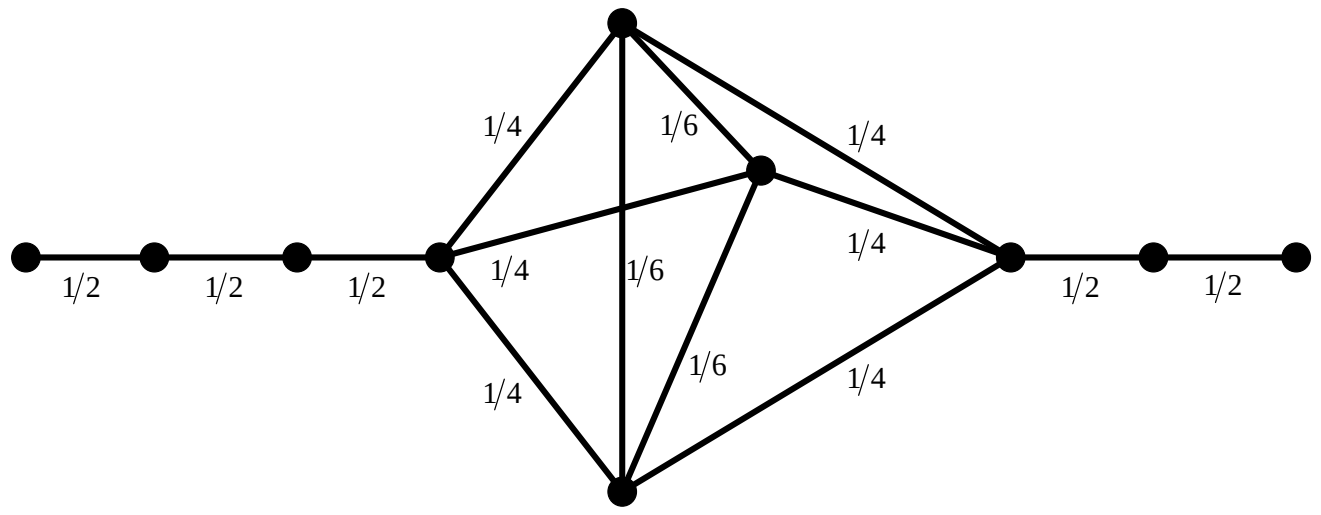

Fig. 8. A semi-complete graph $K_m$ with two path branches $P_{n_1}$ and $P_{n_2}$ for $n_1=2, n_2=3$ and $m=5$.

Dividing the network depicted in Fig. 8 from every node on both path branches splits it into a semi-complete and path branch. The optimal weights of the edges of path branches $P_{n_1}$ and $P_{n_2}$ equal $1/2$ and the optimal weights of the edges connecting inner nodes to each other equal $1/6$ and the edges connecting outer nodes to inner nodes in the semi-complete graph equal $1/4$, respectively. According to subsection III-B, the stratification of the network reduces it to a path graph $P_{n_1+n_2+3}$ with $n_1+n_2+3$ nodes and $(m-3)$ disconnected single nodes. Using the results obtained in [28] and subsection III-B, one can deduce that the semi-complete graph with two path branches has $m-3$ zero eigenvalues and $n_1+n_2+3$ non zero eigenvalues, which are $\cos(k\theta)$ for $k=1,\dots,n_1+n_2+2$ with $\theta=\pi/(n_1+n_2+3)$ and one eigenvalue equal to 1. Thus the *SLEM* equals $\cos(\pi/n_1+n_2+3)$.

### E. Complete Ladder Graph with Two Path branches

As a simple example for the ladder branch introduced in section IV-D we can consider a Ladder graph with $2(m+1)$ nodes connected to two path branches $P_{n_1}$ and $P_{n_2}$ by four bridges as depicted in Fig. 5 for $n_1=2, n_2=3$ and $m=2$.

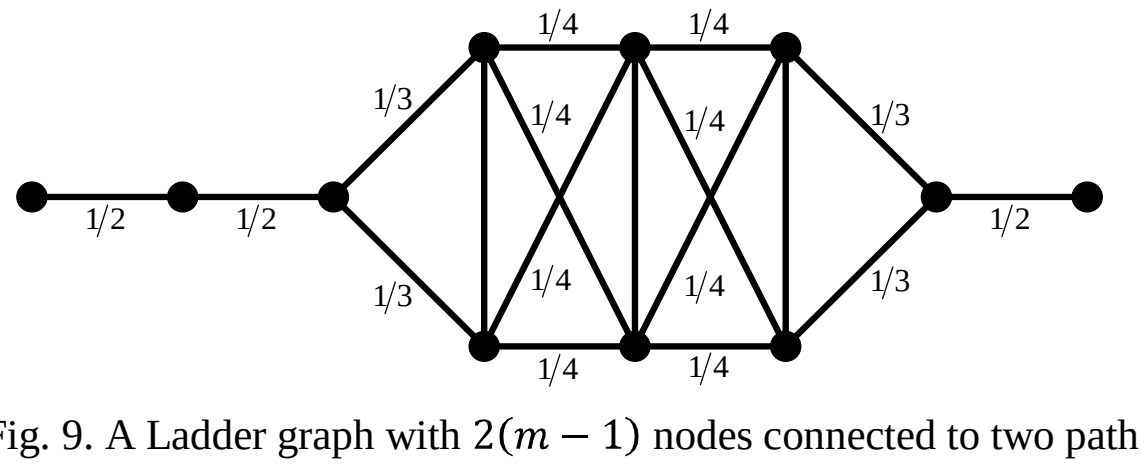

Fig. 9. A Ladder graph with $2(m-1)$ nodes connected to two path branches $P_{n_1}$ and $P_{n_2}$ for $n_1=2, n_2=3$ and $m=2$.

In the network depicted in Fig. 9 each node on one of path branches splits the network into a ladder branch and path branch. The results obtained in section VI-C imply that the optimal weights equal $1/2$ for the edges of path branches ($P_{n_1}$ and $P_{n_2}$) and $1/3$ for four bridges connecting path graphs $P_{n_1}$ and $P_{n_2}$ to complete ladder graph $CL_m$ and $1/4$ for the edges of complete ladder graph as depicted in Fig. 9 for $n_1=2, n_2=3$ and $m=2$.

## VI. Proof of the Results

In this section solution of fastest distributed consensus averaging problem and determination of optimal weights for semi-complete branch introduced in section IV is presented. Due to lack of space for Lollipop, Ladder and palm branches we have only presented the stratification of network's connectivity graph.

### A. Lollipop branch

A Lollipop branch $L_{m,n}$ consists of a path graph $P_n$ and a complete graph $K_m$, which we call them the known part of the network.

We denote the set of nodes of path graph $P_n$ by $\{(1),(2),\dots,(n)\}$ and the nodes of complete graph $K_m$ by $\{(0),(-1),(-2),\dots,(-(m-1))\}$ where $(0)$ is the node connected to path graph $P_n$, (see Fig. 3 for $n=3$, $m=5$).

Automorphism of Lollipop branch is $S_{m-1}$ permutation of nodes of complete graph which are not connected to path graph. Hence according to subsection III-B it has $n+2$ orbits, acting on vertices which are

$$\{(-1),(-2),\dots,(-(m-1))\},\{(0)\},\{(1)\},\{(2)\},\dots,\{(n)\}$$

and $n+2$ class of edge orbits on the known part of network. Thus it suffices to consider just $n+2$ weights $w_{-1}, w_0, w_1, \dots, w_n$ (as labeled in Fig. 3 for $n=3$, $m=5$).

We associate with the node $(i)$, the $|\mathcal{V}|\times 1$ column vector $e_i \in \mathbf{R}_{|\mathcal{V}|}$ (where $|\mathcal{V}|$ is the total number of nodes of network) with 1 in the $i$-th position, and zero elsewhere.
Introducing the new basis $\varphi_i = e_i$, for $i=0,\dots,n-1$ and $\varphi_{-1,\mu} = \frac{1}{\sqrt{m-1}}\sum_{k=0}^{m-2}\omega^{k\mu}e_{-(k+1)}$ for $\mu=0,\dots,m-2$ where $\omega = e^{j\frac{2\pi}{m-1}}$, the weight matrix for the known part of network in the new basis can be defined as

$$\boldsymbol{W} = \begin{bmatrix} (1-w_0-(m-1)w_{-1})\times I_{m-2} & \mathbf{0} \\ \mathbf{0} & \boldsymbol{W'} \end{bmatrix}$$

where $I_{m-2}$ is the identity matrix of dimension $(m-2)$ and $\boldsymbol{W'}$ is as follows:

$$\boldsymbol{W}'_{i,j} = \begin{cases} 1-w_0 \quad \mathbf{for} \quad i=j=1 \\ 1-(m-1)w_0-w_1 \quad \mathbf{for} \quad i=j=2 \\ 1-w_{i-2}-w_{i-1} \quad \mathbf{for} \quad i=j=3,\dots,n+1 \\ \sqrt{m-1}w_0 \quad \mathbf{for} \quad i=j-1=1 \\ \sqrt{m-1}w_0 \quad \mathbf{for} \quad i=j+1=2 \\ w_{i-1} \quad \mathbf{for} \quad i=j-1=2,\dots,n \\ w_{i-2} \quad \mathbf{for} \quad i=j+1=3,\dots,n+1 \end{cases}$$

### B. Semi-Complete Branch

In this section we solve the Fastest Distributed Consensus (FDC) averaging problem for an arbitrary network with a semi-complete branch, using stratification and Semidefinite

Programming (SDP). We consider a network with the undirected associated connectivity graph $\mathcal{G}=(\mathcal{V},\mathcal{E})$, where $\mathcal{V}$ is the set of nodes and $\mathcal{E}$ is the set of edges. The whole branch is called the known part of network and consists of two path links with $n_1$ and $n_2$ edges and one semi-complete graph with $m$ nodes. We denote the set of nodes on the whole branch by $\{(1),(2),\dots,(m+n_1+n_2),\dots\}$ (see Fig. 4 for $n_1=2, n_2=4, m=5$).

Automorphism of semi-complete branch is $S_{m-2}$ permutation of nodes on semi-complete graph which are not connected to path graphs. According to subsection III-B it has $n_1+n_2+3$ class of edge orbits. Thus it suffices to consider just $n_1+n_2+3$ weights namely $w_1,\dots,w_{n_1+n_2+3}$ (as labeled in Fig. 4 for $n_1=2, n_2=4, m=5$). Therefore the weight matrix of semi-complete branch can be written as:

$$\boldsymbol{W}_{i,j}=\begin{cases} w_i \ \textbf{for} \ i=1,\dots,n_1 \quad j=i+1 \\ w_{n_1+1} \ \textbf{for} \ i=n_1+1, \ j=n_1+2,\dots,n_1+m-1 \\ w_{n_1+2} \ \textbf{for} \ i,j=n_1+2,\dots,n_1+m-1 \\ w_{n_1+3} \ \textbf{for} \ i=n_1+2,\dots,n_1+m-1 \quad j=m+n_1 \\ w_{i-m+4} \ \textbf{for} \ i=m+n_1,\dots,m+n_1+n_2-1, \ j=i+1 \\ 1-w_1 \ \textbf{for} \ i=j=1 \\ 1-w_{i-1}-w_i \ \textbf{for} \ i=j=2,\dots,n_1 \\ 1-w_{n_1}-(m-2)w_{n_1+1} \ \textbf{for} \ i=j=n_1+1 \\ 1-w_{n_1+1}-w_{n_1+3}-(m-3)w_{n_1+2} \ \textbf{for} \ i=j=n_1+2,\dots,n_1+m-1 \\ 1-w_{n_1+4}-(m-2)w_{n_1+3} \ \textbf{for} \ i=j=m+n_1 \\ 1-w_{i-m+3}-w_{i-m+4} \ \textbf{for} \ i=j=m+n_1+1,\dots,m+n_1+n_2-1 \\ 0 \ \textbf{Otherwise} \end{cases}$$

We assign with node $(i)$ the $|\mathcal{V}|\times 1$ column vector $e_i$ (where $|\mathcal{V}|$ is the total number of nodes on network) with 1 in the $i$-th position and zero elsewhere, then the edge $\{i,j\}\in\mathcal{E}$ can be written as $e_i e_j^T$. In the new basis defined as

$$\varphi_i=\begin{cases} \dfrac{1}{\sqrt{m-2}}\sum_{k=0}^{m-3}\omega^{ki}e_{n_1+2+k} \ \ \textbf{for} \ i=1,\dots,m-3 \\ e_{i-m+3} \ \textbf{for} \ i=m-2,\dots,n_1+m-2 \\ \dfrac{1}{\sqrt{m-2}}\sum_{k=0}^{m-3}e_{n_1+2+k} \ \ \textbf{for} \ i=n_1+m-1 \\ e_i \ \textbf{for} \ i=n_1+m,\dots,n_1+n_2+m \end{cases}$$

with $\omega=exp\left(j\frac{2\pi}{m-2}\right)$, the weight matrix $\boldsymbol{W}$ for semi-complete branch takes the form as

$$\boldsymbol{W_{i,j}}=\begin{cases} 1-w_{n_1+1}-w_{n_1+3}-(m-2)w_{n_1+2} \ \textbf{for} \ i=j=1,\dots,m-3 \\ 1-w_1 \ \textbf{for} \ i=j=m-2 \\ 1-w_{i-m+2}-w_{i-m+3} \ \textbf{for} \ i=j=m-1,\dots,m+n_1-3 \\ 1-w_{n_1}-(m-2)w_{n_1+1} \ \textbf{for} \ i=j=m+n_1-2 \\ 1-w_{n_1+1}-w_{n_1+3} \ \textbf{for} \ i=j=m+n_1-1 \\ 1-w_{n_1+4}-(m-2)w_{n_1+3} \ \textbf{for} \ i=j=m+n_1 \\ 1-w_{i-m+3}-w_{i-m+4} \ \textbf{for} \ i=j=m+n_1+1,\dots,m+n_1+n_2-1 \\ w_{i-m+3} \ \textbf{for} \ i=m-2,\dots,m+n_1-3, \ j=i+1 \\ \sqrt{m-2}w_{n_1+1} \ \textbf{for} \ i=m+n_1-2, \ j=i+1 \\ \sqrt{m-2}w_{n_1+3} \ \textbf{for} \ i=m+n_1-1, \ j=i+1 \\ w_{n_1+4} \ \textbf{for} \ i=m+n_1, \ j=i+1 \\ w_{i-m+4} \ \textbf{for} \ i=m+n_1+1,\dots,m+n_1+n_2-1, \ j=i+1 \\ 0 \ \textbf{Otherwise} \end{cases}$$

The off diagonal elements of first $(m-3)$ rows and columns of $\boldsymbol{W}$ are zero and the diagonal entries equal $\left(1-w_{n_1+1}-w_{n_1+3}-(m-2)w_{n_1+2}\right)$ which are the eigenvalues as well and by considering the fact that *SLEM* is the second largest eigenvalue in magnitude we should have

$$\left|1-w_{n_1+1}-w_{n_1+3}-(m-2)w_{n_1+2}\right|\le SLEM$$

which is definitely satisfied by

$$w_{n_1+2}=\frac{1-w_{n_1+1}-w_{n_1+3}}{m-2} \tag{3}$$

Based on subsection III-A, one can express FDC problem for semi-complete branch in the form of semidefinite programming as:

$$\begin{aligned} \min \quad & s \\ s.t. \quad & -sI\preccurlyeq \boldsymbol{W}-\boldsymbol{v}\boldsymbol{v}^T\preccurlyeq sI \end{aligned} \tag{4}$$

where $\boldsymbol{v}$ is a $|\mathcal{V}|\times 1$ column vector defined as:

$$\boldsymbol{v}_i=\frac{1}{\sqrt{|\mathcal{V}|}}\times\begin{cases} 0 & \text{for} \ i=1,\dots,m-3 \\ \sqrt{m-1} & \text{for} \ i=m+n_1-1 \\ 1 & \textbf{Otherwise} \end{cases}$$

which is eigenvector of $\boldsymbol{W}$ corresponding to the eigenvalue one. The weight matrix $\boldsymbol{W}$ can be written as

$$\begin{aligned}\boldsymbol{W}=I_{|\mathcal{V}|}-(m-2)w_{n_1+2}\begin{bmatrix} I_{m-3} & 0 & 0 \\ 0 & 0 & 0 \\ 0 & 0 & 0 \end{bmatrix}-w_{n_1+1}\begin{bmatrix} I_{m-3} & 0 & 0 \\ 0 & \boldsymbol{\alpha}_{n_1+1}\boldsymbol{\alpha}_{n_1+1}^T & 0 \\ 0 & 0 & 0 \end{bmatrix} \\ -w_{n_1+3}\begin{bmatrix} I_{m-3} & 0 & 0 \\ 0 & \boldsymbol{\alpha}_{n_1+3}\boldsymbol{\alpha}_{n_1+3}^T & 0 \\ 0 & 0 & 0 \end{bmatrix}-\sum_{\substack{i=1 \\ i\neq n_1+1,n_1+2,n_1+3}}^{n_1+n_2+3} w_i\begin{bmatrix} 0 & 0 & 0 \\ 0 & \boldsymbol{\alpha}_i\boldsymbol{\alpha}_i^T & 0 \\ 0 & 0 & 0 \end{bmatrix}-\boldsymbol{W}'\end{aligned} \tag{5}$$

where $\boldsymbol{W'}$ includes the weights on the unknown part of network and $\boldsymbol{\alpha}_i$ for $i=1,\dots,n_1+n_2+3, i\neq n_1+2$ are $(n_1+n_2+3)\times 1$ column vectors defined as

$$\boldsymbol{\alpha}_i(j)=\begin{cases} 1 \ \textbf{for} \ j=i \\ -1 \ \textbf{for} \ j=i+1 \\ 0 \ \textbf{Otherwise} \end{cases} \ \textbf{for} \ i=1,\dots,n_1$$

$$\boldsymbol{\alpha}_{n_1+1}(j)=\begin{cases} \sqrt{m-2} \ \textbf{for} \ j=n_1+1 \\ -1 \ \textbf{for} \ j=n_1+2 \\ 0 \ \textbf{Otherwise} \end{cases}$$

$$\boldsymbol{\alpha}_i(j)=\begin{cases} 1 \ \textbf{for} \ j=i-1 \\ -1 \ \textbf{for} \ j=i \\ 0 \ \textbf{Otherwise} \end{cases} \ \textbf{for} \ i=n_1+4,\dots,n_1+n_2+3$$

$$\boldsymbol{\alpha}_{n_1+3}(j)=\begin{cases} 1 \ \textbf{for} \ j=n_1+2 \\ -\sqrt{m-2} \ \textbf{for} \ j=n_1+3 \\ 0 \ \textbf{Otherwise} \end{cases}$$

In order to formulate problem (4) in the form of standard semidefinite programming described in section II-C, we define $F_i, c_i$ and $x$ as below:

$$F_0=-\boldsymbol{\sigma}_z\otimes\left(I_{|\mathcal{V}|}-\boldsymbol{v}\boldsymbol{v}^T\right), \quad F_1=I_{2|\mathcal{V}|},$$

$$F_i=\boldsymbol{\sigma}_z\otimes\begin{bmatrix} 0 & 0 & 0 \\ 0 & \boldsymbol{\alpha}_{i-1}\boldsymbol{\alpha}_{i-1}^T & 0 \\ 0 & 0 & 0 \end{bmatrix} \ \textbf{for} \ \begin{matrix} i=2,\dots,n_1+n_2+4, \\ i\neq n_1+2,n_1+3,n_1+4 \end{matrix}$$

$$F_{n_1+2}=\boldsymbol{\sigma}_z\otimes\begin{bmatrix} I_{m-3} & 0 & 0 \\ 0 & \boldsymbol{\alpha}_{n_1+1}\boldsymbol{\alpha}_{n_1+1}^T & 0 \\ 0 & 0 & 0 \end{bmatrix}$$

$$F_{n_1+3}=\boldsymbol{\sigma}_z\otimes(m-2)\begin{bmatrix} I_{m-3} & 0 & 0 \\ 0 & 0 & 0 \\ 0 & 0 & 0 \end{bmatrix}$$

$$F_{n_1+4}=\boldsymbol{\sigma}_z\otimes\begin{bmatrix} I_{m-3} & 0 & 0 \\ 0 & \boldsymbol{\alpha}_{n_1+3}\boldsymbol{\alpha}_{n_1+3}^T & 0 \\ 0 & 0 & 0 \end{bmatrix}$$

$$c_1=1, \quad c_i=0, \ i=2,\dots$$

$$x^T = [s, w_1, \dots, w_{n_1+n_2+1}, \dots]$$

where $\boldsymbol{\sigma}_z = \begin{bmatrix} 1 & 0 \\ 0 & -1 \end{bmatrix}$. In the dual case we choose the dual variable $Z \geq 0$ as $Z = \begin{bmatrix} z_1 \\ z_2 \end{bmatrix} \cdot [z_1^T \quad z_2^T]$ where $z_1$, and $z_2$ are column vectors each with $|\mathcal{V}|$ elements. It is obvious that $Z$ is positive definite. Resolving each of $z_1$ and $z_2$ into three orthogonal vectors as

$$z_1 = \begin{bmatrix} z_{1,1} \\ z_{1,2} \\ z_{1,3} \end{bmatrix}, z_2 = \begin{bmatrix} z_{2,1} \\ z_{2,2} \\ z_{2,3} \end{bmatrix}$$

where $z_{1,1}, z_{1,2}$ and $z_{1,3}$ are $(m-3) \times 1$, $(n_1+n_2+3) \times 1$ and $(|\mathcal{V}| - n_1 - n_2 - m) \times 1$ column vectors respectively and the same holds for $z_{2,1}, z_{2,2}$ and $z_{2,3}$.

From the complementary slackness condition (2) we have

$$(sI - \boldsymbol{W} + \boldsymbol{v}\boldsymbol{v}^T) z_1 = 0, \;\; (sI + \boldsymbol{W} - \boldsymbol{v}\boldsymbol{v}^T) z_2 = 0 \qquad (6)$$

Multiplying both sides of equations (6) by $\boldsymbol{v}\boldsymbol{v}^T$ we have $s(\boldsymbol{v}\boldsymbol{v}^T z_1) = 0$ and $s(\boldsymbol{v}\boldsymbol{v}^T z_2) = 0$ which implies that $\boldsymbol{v}^T z_1 = 0$ and $\boldsymbol{v}^T z_2 = 0$. Consequently (6) reduces to

$$(sI - \boldsymbol{W}) z_1 = 0, \;\; (sI + \boldsymbol{W}) z_2 = 0 \qquad (7)$$

Using the constraints $Tr[F_i Z] = c_i$ we have

$$z_1^T z_1 + z_2^T z_2 = 1$$

$$z_{1,1}^T z_{1,1} - z_{2,1}^T z_{2,1} = 0 \qquad (8\text{-a})$$

$$z_{1,1}^T z_{1,1} - z_{2,1}^T z_{2,1} + \left(\boldsymbol{\alpha}_{n_1+1}^T z_{1,2}\right)^2 - \left(\boldsymbol{\alpha}_{n_1+1}^T z_{2,2}\right)^2 = 0 \qquad (8\text{-b})$$

$$z_{1,1}^T z_{1,1} - z_{2,1}^T z_{2,1} + \left(\boldsymbol{\alpha}_{n_1+3}^T z_{1,2}\right)^2 - \left(\boldsymbol{\alpha}_{n_1+3}^T z_{2,2}\right)^2 = 0 \qquad (8\text{-c})$$

$$\left(\boldsymbol{\alpha}_i^T z_{1,2}\right)^2 - \left(\boldsymbol{\alpha}_i^T z_{2,2}\right)^2 = 0 \;\; \textbf{for} \;\; \begin{matrix} i = 1, \dots, n_1+n_2+2, \\ i \neq n_1+1, n_1+2, n_1+3 \end{matrix} \qquad (8\text{-d})$$

To have the strong duality we set $c^T x + Tr[F_0 Z] = 0$, hence we have $z_1^T z_1 - z_2^T z_2 = s$. Considering the linear independence of $\boldsymbol{\alpha}_i$ for $i = 1, \dots, n_1+n_2+3$, we can expand $z_{1,2}$ and $z_{2,2}$ in terms of $\boldsymbol{\alpha}_i$ as

$$z_{1,2} = \sum_{\substack{i=1 \\ i \neq n_1+2}}^{n_1+n_2+3} a_i \boldsymbol{\alpha}_i, \;\; z_{2,2} = \sum_{\substack{i=1 \\ i \neq n_1+2}}^{n_1+n_2+3} a_i' \boldsymbol{\alpha}_i \qquad (9)$$

with the coordinates $a_i$ and $a_i'$, $i = 1, \dots, n_1+n_2+3$, $i \neq n_1+2$ to be determined.

Using (5) and the expansions (9), from equalizing the coordinates of $\boldsymbol{\alpha}_i$ for $i = 1, \dots, n_1+n_2+3$, $i \neq n_1+2$ in the slackness conditions (7), we have

$$w_i \boldsymbol{\alpha}_i^T z_{1,2} = (-s+1) a_i, \;\; w_i \boldsymbol{\alpha}_i^T z_{2,2} = (s+1) a_i' \qquad (10)$$

where (10) hold for $i = 1, \dots, n_1+n_2+2$, $i \neq n_1+2$. Considering (8-b), (8-c), (8-d) and (8-e) we obtain $(s-1)^2 a_i^2 = (s+1)^2 a_i'^2$, for $i = 1, \dots, n_1+n_2+2$, $i \neq n_1+2$, or equivalently

$$\left(a_i / a_j\right)^2 = \left(a_i' / a_j'\right)^2 \qquad (11)$$

for $\forall i, j = 1, \dots, n_1+n_2+2$, $i, j \neq n_1+2$ and for $\boldsymbol{\alpha}_i^T z_{1,2}$ and $\boldsymbol{\alpha}_i'^T z_{2,2}$, we have

$$\boldsymbol{\alpha}_i^T z_{1,2} = \sum_{\substack{j=1 \\ j \neq n_1+2}}^{n_1+n_2+3} a_j G_{i,j}, \;\; \boldsymbol{\alpha}_i^T z_{2,2} = \sum_{\substack{j=1 \\ j \neq n_1+2}}^{n_1+n_2+3} a_j' G_{i,j} \qquad (12)$$

where $G$ is the Gram matrices, defined as $G_{i,j} = \boldsymbol{\alpha}_i^T \boldsymbol{\alpha}_j$, or equivalently

$$G = \begin{bmatrix}
2 & -1 & 0 & & & & & & & \\
-1 & 2 & \ddots & \ddots & & & & & & \\
0 & \ddots & \ddots & -1 & 0 & & & & & \\
 & \ddots & -1 & 2 & -\sqrt{m-1} & 0 & & & & \\
 & & 0 & -\sqrt{m-1} & m-1 & -1 & 0 & & & \\
 & & & 0 & -1 & m-1 & -\sqrt{m-1} & 0 & & \\
 & & & & 0 & -\sqrt{m-1} & 2 & -1 & \ddots & \\
 & & & & & 0 & -1 & \ddots & \ddots & 0 \\
 & & & & & & \ddots & \ddots & 2 & -1 \\
 & & & & & & & 0 & -1 & 2
\end{bmatrix}$$

Substituting (12) in (10) we have

$$(-s+1-2w_1) a_1 = -w_1 a_2 \qquad (13\text{-a})$$

$$(-s+1-2w_i) a_i = -w_i (a_{i-1} + a_{i+1}) \qquad (13\text{-b})$$

$$\left(-s+1-2w_{n_1}\right) a_{n_1} = -w_{n_1}\left(a_{n_1-1} + \hat{a}_{n_1+1}\right) \qquad (13\text{-c})$$

$$\left(-s+1-(m-1)w_{n_1+1}\right) \hat{a}_{n_1+1} = -w_{n_1+1}\left((m-2) a_{n_1} + \hat{a}_{n_1+3}\right) \qquad (13\text{-d})$$

$$\left(-s+1-(m-1)w_{n_1+3}\right) \hat{a}_{n_1+3} = -w_{n_1+3}\left(\hat{a}_{n_1+1} - (m-2) a_{n_1+4}\right) \qquad (13\text{-e})$$

$$\left(-s+1-2w_{n_1+4}\right) a_{n_1+4} = -w_{n_1+4}\left(\hat{a}_{n_1+3} + a_{n_1+5}\right) \qquad (13\text{-f})$$

and

$$(s+1-2w_1) a_1' = -w_1 a_2' \qquad (14\text{-a})$$

$$(s+1-2w_i) a_i' = -w_i (a_{i-1}' + a_{i+1}') \qquad (14\text{-b})$$

$$\left(s+1-2w_{n_1}\right) a_{n_1}' = -w_{n_1}\left(a_{n_1-1}' + \hat{a}_{n_1+1}'\right) \qquad (14\text{-c})$$

$$\left(s+1-(m-1)w_{n_1+1}\right) \hat{a}_{n_1+1}' = -w_{n_1+1}\left((m-2) a_{n_1}' + \hat{a}_{n_1+3}'\right) \qquad (14\text{-d})$$

$$\left(s+1-(m-1)w_{n_1+3}\right) \hat{a}_{n_1+3}' = -w_1\left(\hat{a}_{n_1+1}' - (m-2) a_{n_1+4}'\right) \qquad (14\text{-e})$$

$$\left(s+1-2w_{n_1+4}\right) a_{n_1+4}' = -w_{n_1+4}\left(\hat{a}_{n_1+3}' + a_{n_1+5}'\right) \qquad (14\text{-f})$$

where $\hat{a}_{n_1+1} = \sqrt{m-2}\, a_{n_1+1}$, $\hat{a}_{n_1+1}' = \sqrt{m-2}\, a_{n_1+1}'$, $\hat{a}_{n_1+3} = \sqrt{m-2}\, a_{n_1+3}$, $\hat{a}_{n_1+3}' = \sqrt{m-2}\, a_{n_1+3}'$ and (13-b) and (14-b) hold for $i = 2, \dots, n_1+n_2+2$, $i \neq n_1, n_1+1, n_1+2, n_1+3, n_1+4$.

Now we can determine the optimal weights in an inductive manner as follows:

In the first stage, from comparing equations (13-a) and (14-a) and considering the relation (11), we can conclude that $(-s+1-2w_1)^2 = (s+1-2w_1)^2$, which results in $w_1 = 1/2$ and $s = 0$, where the latter is not acceptable. Assuming $s = \cos(\theta)$ and substituting $w_1 = 1/2$ in (13-a) and (14-a), we have

$$a_2 = \frac{\sin(2\theta)}{\sin(\theta)} a_1, \;\; a_2' = \frac{\sin\left(2(\pi-\theta)\right)}{\sin(\pi-\theta)} a_1'$$

Continuing the above procedure inductively, up to $i-1$ stages, and assuming

$$a_j = \frac{\sin(j\theta)}{\sin(\theta)} a_1, \qquad a_j' = \frac{\sin\left(j(\pi-\theta)\right)}{\sin(\pi-\theta)} a_1' \qquad \forall j \leq i$$

in the $i$-th stage, we get the following equations from comparison of equations (13-b) and (14-b),

$$\left((-s+1-2w_i)\frac{\sin(i\theta)}{\sin(\theta)} + w_i \frac{\sin\left((i+1)\theta\right)}{\sin(\theta)}\right) a_1 = -w_i a_{i+1} \qquad (15\text{-a})$$

$$\left((s+1-2w_i)\frac{\sin\left(i(\pi-\theta)\right)}{\sin(\pi-\theta)} + w_i \frac{\sin\left((i+1)(\pi-\theta)\right)}{\sin(\pi-\theta)}\right) a_1' = -w_i a_{i+1}' \qquad (15\text{-b})$$

while considering relation (11) we can conclude that

$$\left((-s+1-2w_i)\sin(i\theta) + w_i \sin\left((i+1)\theta\right)\right)^2 =$$

$$\left((s+1-2w_i)\sin\left(i(\pi-\theta)\right) + w_i \sin\left((i+1)(\pi-\theta)\right)\right)^2$$

which results in

$$w_i = 1/2 \tag{16}$$

Substituting $w_i = 1/2$ in (15), we have

$$a_{i+1} = \frac{\sin\left((i+1)\theta\right)}{\sin(\theta)} a_1,\ \ a'_{i+1} = \frac{\sin\left((i+1)(\pi-\theta)\right)}{\sin(\pi-\theta)} a'_1 \tag{17}$$

where (16) and (17) are true for $i = 2, \dots, n_1 - 1$ and in the $n_1$-th stage, in the same way as in previous stages from equations (13-c) and (14-c) we have

$$w_{n_1} = 1/2$$

$$\hat{a}_{n_1+1} = \frac{sin\left((n_1+1)\theta\right)}{sin(\theta)} a_1,\ \ \hat{a}'_{n_1+1} = \frac{sin\left((n_1+1)(\pi-\theta)\right)}{sin(\pi-\theta)} a'_1$$

In the $(n_1+1)$-th stage, from equations (13-d) and (14-d) and using relation (11), we can conclude that

$$\left(\left(-s+1-(m-1)w_{n_1+1}\right)\sin\left((n_1+1)\theta\right) + w_{n_1+1}(m-2)\sin(n_1\theta)\right)^2 = \left(\left(s+1-(m-1)w_{n_1+1}\right)sin\left((n_1+1)\hat{\theta}\right) + w_{n_1+1}(m-2)sin\left(n_1\hat{\theta}\right)\right)^2$$

where $\hat{\theta} = \pi - \theta$. The relation above results in $w_{n_1+1} = 1/m - 1$ and $s = 0$, where the latter is not acceptable and

$$\hat{a}_{n_1+3} = \frac{sin\left((n_1+2)\theta\right)}{sin(\theta)} a_1,\ \ \hat{a}'_{n_1+3} = \frac{sin\left((n_1+2)(\pi-\theta)\right)}{sin(\pi-\theta)} a'_1.$$

In the $(n_1+2)$-th stage, from equations (13-e) and (14-e) and using relation (11), we can conclude that

$$\left(\left(-s+1-(m-1)w_{n_1+3}\right)\sin\left((n_1+2)\theta\right) + w_{n_1+3}\sin\left((n_1+1)\theta\right)\right)^2 = \left(\left(s+1-(m-1)w_{n_1+3}\right)sin\left((n_1+2)\hat{\theta}\right) + w_{n_1+3}sin\left((n_1+1)\hat{\theta}\right)\right)^2$$

which results in $w_{n_1+3} = 1/m - 1$ and $s = 0$, where the latter is not acceptable and doing the same inductive procedure as explained in previous stages, from equations (13-b) and (14-b) we have

$$w_i = 1/2 \tag{18}$$

$$a_{i+1} = \frac{\sin(i\theta)}{\sin(\theta)} a_1,\ \ a'_{i+1} = \frac{\sin\left(i(\pi-\theta)\right)}{\sin(\pi-\theta)} a'_1 \tag{19}$$

where (18) and (19) are true for $i = n_1 + 4, \dots, n_1 + n_2 + 2$. Using (3) we have

$$w_{n_1+2} = \frac{m-3}{(m-1)(m-2)}$$

## C. Ladder Branch

A ladder branch consists of two path graphs with $n_1$ and $n_2$ edges connected by four bridges to a complete ladder graph with $2(m+1)$ nodes. We call the whole branch, the known part of network and denote the set of nodes on the whole branch by $\{(1), (2), \dots, (n_1+n_2+2m+4), \dots\}$ (see Fig. 5).

Automorphism of ladder branch is $S_2$ permutation of nodes on the same level of complete ladder graph. According to subsection III-B it has $n_1 + n_2 + 2$ and $2m+1$ class of edge orbits on path graphs and the complete ladder graph, respectively. Thus it suffices to consider just $n_1 + n_2 + 2m + 3$ weights namely $w_1, \dots, w_{n_1+n_2+m+2}, w'_1, \dots, w'_{m+1}$ (as labeled in Fig. 5). We assign with node $(i)$ the $|\mathcal{V}| \times 1$ column vector $e_i$ (where $|\mathcal{V}|$ is the total number of nodes of network) with 1 in the $i$-th position and zero elsewhere, then the edge $\{i,j\} \in \mathcal{E}$ can be written as $e_i e_j^T$. In the new basis defined as

$$\varphi_i = \begin{cases} \frac{1}{\sqrt{2}}\left(e_{2i+n_1} - e_{2i+n_1+1}\right) & \textbf{for}\ \ i = 1, \dots, m+1 \\ e_{i-m-1} & \textbf{for}\ \ i = m+2, \dots, n_1+m+2 \\ \frac{1}{\sqrt{2}}\left(e_{2i-n_1-2m-4} + e_{2i-n_1-2m-3}\right) & \textbf{for}\ \ i = n_1+m+3, \dots, n_1+2m+3 \\ e_i & \textbf{for}\ \ i = n_1+2m+4, \dots, n_1+2m+n_2+3 \end{cases}$$

the weight matrix of ladder branch can be defined as:

$$\boldsymbol{W_{i,j}} = \begin{cases} 1 - 2w_{n_1+2} - w_{n_1+1} - 2w'_1 & \textbf{for}\ \ i = j = 1 \\ 1 - 2w_{i+n_1+1} - 2w_{i+n_1} - 2w'_i & \textbf{for}\ \ i = j = 2, \dots, m \\ 1 - 2w_{n_1+m+1} - w_{n_1+m+2} - 2w'_{m+1} & \textbf{for}\ \ i = j = m+1 \\ 1 - w_1 & \textbf{for}\ \ i = j = m+2 \\ w_{i-m-1} & \textbf{for}\ \ i = m+2, \dots, n_1+m+1,\ \ j = i+1 \\ 1 - w_{i-m-2} - w_{i-m-1} & \textbf{for}\ \ i = j = m+2, \dots, n_1+m+1 \\ 1 - w_{n_1} - 2w_{n_1+1} & \textbf{for}\ \ i = j = n_1+m+2 \\ \sqrt{2}w_{n_1+1} & \textbf{for}\ \ i = n_1+m+2,\ \ j = i+1 \\ 2w_{i-m-1} & \textbf{for}\ \ i = n_1+m+3, \dots, n_1+2m+2,\ \ j = i+1 \\ \sqrt{2}w_{n_1+m+2} & \textbf{for}\ \ i = n_1+2m+3,\ \ j = i+1 \\ 1 - 2w_{n_1+2} - w_{n_1+1} & \textbf{for}\ \ i = j = n_1+m+3 \\ 1 - 2w_{i-m-1} - 2w_{i-m-2} & \textbf{for}\ \ i = j = n_1+m+4, \dots, n_1+2m+2 \\ 1 - 2w_{n_1+m+1} - w_{n_1+m+2} & \textbf{for}\ \ i = j = n_1+2m+3 \\ w_{i-m-1} & \textbf{for}\ \ i = n_1+2m+4, \dots, n_1+2m+n_2+3 \\ 1 - 2w_{n_1+m+2} - w_{n_1+m+3} & \textbf{for}\ \ i = j = n_1+2m+4 \\ 1 - w_{i-m-2} - w_{i-m-1} & \textbf{for}\ \ i = j = n_1+2m+5, \dots, n_1+2m+n_1+3 \\ 0 & \textbf{Otherwise} \end{cases}$$

The off diagonal elements of first $(m+1)$ rows and columns of $\boldsymbol{W}$ are zero and the diagonal entries equal $\left(1 - 2w_{n_1+2} - w_{n_1+1} - 2w'_1\right)$, $\left(1 - 2w_{i+n_1+1} - 2w_{i+n_1} - 2w'_i\right)$ for $i = 2, \dots, m$ and $\left(1 - 2w_{n_1+m+1} - w_{n_1+m+2} - 2w'_{m+1}\right)$ which are the eigenvalues as well. Considering the fact that *SLEM* is the second largest eigenvalue in magnitude, we can conclude that the *SLEM* of network is independent of $w'_i$ for $i = 1, \dots, m+1$ as long as these weights satisfy the following relations.

$$0 \le \left|1 - 2w_{n_1+2} - w_{n_1+1} - 2w'_1\right| \le SLEM \tag{20-a}$$

$$0 \le \left|1 - 2w_{i+n_1+1} - 2w_{i+n_1} - 2w'_i\right| \le SLEM,\ \ \textbf{for}\ \ i = 2, \dots, m \tag{20-b}$$

$$0 \le \left|1 - 2w_{n_1+m+1} - w_{n_1+m+2} - 2w'_{m+1}\right| \le SLEM \tag{20-c}$$

## D. Palm branch

A palm branch of order $(m,k)$ consists of a path graph $P_m$ and a star graph with $k$ branches of length one, which we call them the known part of the network.

We denote the set of nodes of path graph $P_m$ by $\{(1), (2), \dots, (m)\}$ and the nodes of star graph by $\{(0), (-1), (-2), \dots, (-k)\}$. $(0)$ is the central node of star graph connected to path graph $P_m$, (see Fig. 6 for $m = 3$, $k = 4$).

Automorphism of Lollipop branch is $S_k$ permutation of branches of star graph which are not connected to path graph via central node. Hence according to subsection III-B it has $m+1$ orbits, acting on vertices which are

$$\{(-1), (-2), \dots, (-k)\}, \{(0)\}, \{(1)\}, \{(2)\}, \dots, \{(m)\}$$

and $m+1$ class of edge orbits on the known part of network. Thus it suffices to consider just $m+1$ weights $w_0, w_1, \dots, w_m$ (as labeled in Fig. 6 for $m=3$, $k=4$).

We associate with the node $(i)$, the $|\mathcal{V}| \times 1$ column vector $e_i \in \mathbf{R}_{|\mathcal{V}|}$ (where $|\mathcal{V}|$ is the total number of nodes of network) with 1 in the $i$-th position, and zero elsewhere.

Introducing the new basis $\varphi_i = e_i$, for $i = 0, \dots, m$ and $\varphi_{-1,\mu} = \frac{1}{\sqrt{k}} \sum_{q=1}^{k} \omega^{(q-1)\mu} e_{-q}$ for $\mu = 0, \dots, k-1$ where $\omega = e^{j\frac{2\pi}{k}}$, the weight matrix for the known part of network in the new basis can be defined as

$$\boldsymbol{W} = \begin{bmatrix} (1-w_0) \times I_{k-1} & \mathbf{0} \\ \mathbf{0} & \boldsymbol{W}' \end{bmatrix}$$

where $I_{m-2}$ is the identity matrix of dimension $(m-2)$ and $\boldsymbol{W}'$ is as follows:

$$\boldsymbol{W}'_{i,j} = \begin{cases} 1 - w_0 \quad \textbf{for} \quad i = j = 1 \\ 1 - kw_0 - w_1 \quad \textbf{for} \quad i = j = 2 \\ 1 - w_{i-2} - w_{i-1} \quad \textbf{for} \quad i = j = 3, \dots, m+2 \\ \sqrt{k} w_0 \quad \textbf{for} \quad i = j - 1 = 1 \\ \sqrt{k} w_0 \quad \textbf{for} \quad i = j + 1 = 2 \\ w_{i-1} \quad \textbf{for} \quad i = j - 1 = 2, \dots, m+1 \\ w_{i-2} \quad \textbf{for} \quad i = j + 1 = 3, \dots, m+1 \end{cases}$$

## VII. Convergence Rates of Branches

In this section we aim to compare five branches introduced in section IV in terms of asymptotic and per step convergence rates. Also we have compared the obtained optimal weights with other common weighting methods, namely maximum degree [5], Metropolis-Hasting [12] and best constant [2] weighting methods by evaluating SLEM and comparing convergence time improvements. For this purpose we consider a network with symmetric star topology where 8 similar branches are connected to a central node. In table 1 SLEM of this network for different types of branches and weighting methods is presented. As it is obvious from table 1 optimal weights result in smaller value for SLEM.

| | Optimal | Max Degree | Metropolis-Hasting | Best Constant |
|---|---|---|---|---|
| Path branch of length 10 | 0.99138 | 0.9972 | 0.99468 | 0.99512 |
| Lollipop Branch of order (5,5) | 0.99091 | 0.99691 | 0.9947 | 0.994616 |
| Semi-Complete Branch of order (5,2,3) | 0.98939 | 0.99587 | 0.99412 | 0.9928 |
| Ladder branch of order (1,2,2) | 0.98947 | 0.99593 | 0.99396 | 0.9929 |
| Palm branch of order (4,5) | 0.99043 | 0.99662 | 0.9947 | 0.99411 |

Table 1. SLEM of a symmetric star network with 8 similar branches and 4 different weighting methods.

In Fig. 10 the total error in terms of number of iterations over a symmetric star network with 8 path branches of length 10 is presented. The weighting methods, used for achieving the results of Fig. 10 are optimal weights (given in section IV), Maximum degree, Metropolis-Hasting and Best constant weighting methods (as defined in Appendix A). We define the total error as the Euclidean distance of vector of node values $x(t)$ from the mean of vector of initial node values $(1/n) \sum_{i=1}^{n} x_i(0)$.

In figures 11, 12, 13 and 14 we have had the same comparison as in Fig. 10 but for other types of branches introduced in section IV. In Fig. 11 Lollipop branches of order $(5,5)$, in Fig. 12 Semi-Complete branches of order $(5,2,3)$, in Fig. 13 Ladder branches of order $(1,2,2)$ and in Fig. 14 palm branches of order $(4,5)$ are used as the star networks branches.

As it is obvious from Figures 10, 11, 12, 13 and 14 at first few iterations Metropolis weights has better mixing rate per step compared to other weights but after a few iterations optimal weights achieve better performance than other three weighting methods because of smaller SLEM value. It should be mentioned that the results depicted in Figures 10, 11, 12, 13 and 14 are in logarithmic scale and generated based on 10000 trials (a different random initial node values is generated for each trial).

Each one of the branches considered in table 1 and figures 10, 11, 12, 13 and 14 has ten nodes, but different number of edges. Including the edge which is connecting the branch to the central node, path branch of length 10 has 10 edges, Semi-complete branch of order $(5,2,3)$ has 15 edges, Lollipop branch of order $(5,5)$ has 16 edges, Ladder branch of order $(1,2,2)$ has 15 edges and palm branch of order $(4,5)$ has 10 edges. According to table 1 for a fixed number of nodes in each branch by choosing semi-complete topology and its corresponding optimal weights (given in section IV), one can achieve the fastest asymptotic mixing rate. Not to mention that faster mixing rate comes with the cost of more edges and connections. palm branch has the minimum number of edges required for the network to remain connected and still mixes faster than path and Lollipop branches.

To compare these branches in terms of per step convergence rate, in Fig. 15 total error in terms of number of iterations over a symmetric star network with 8 identical branches is presented. The type of branches considered for the results of Fig. 15 are the same as in table 1.

From Fig. 15 it is obvious that Semi-complete and Ladder branches mix faster compared to path, Lollipop and palm branches.

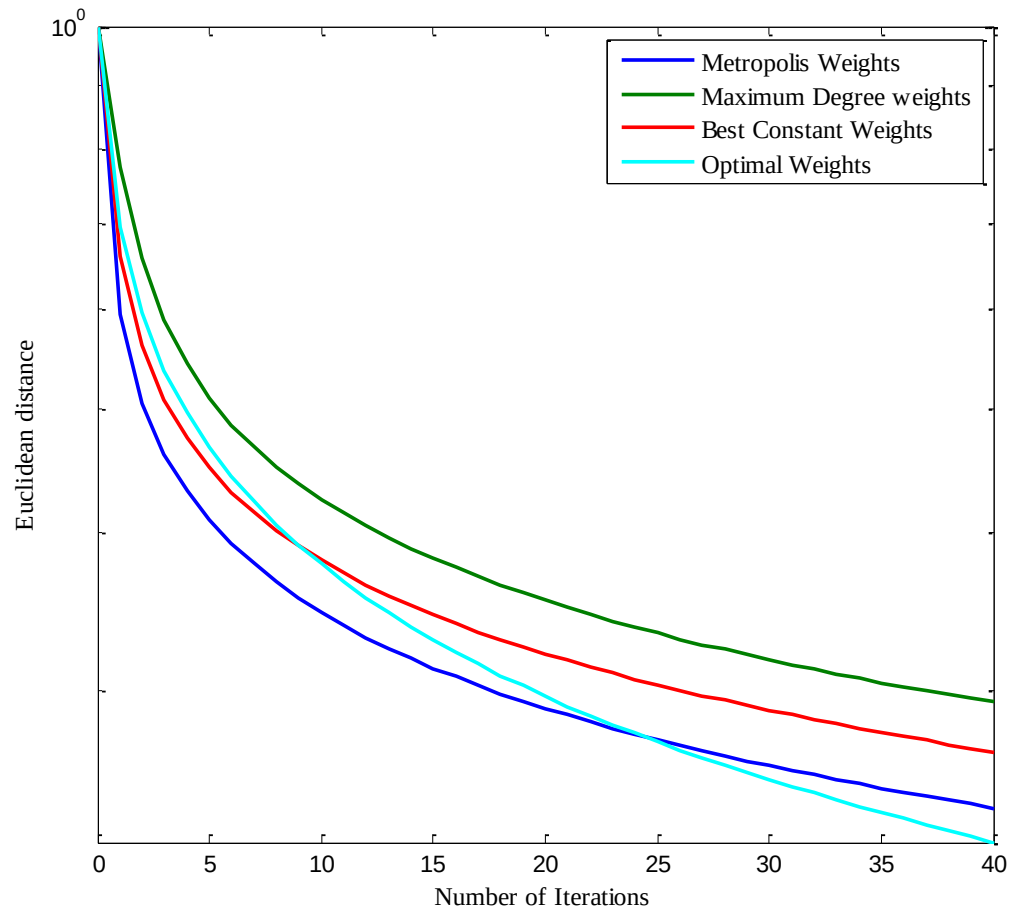

Fig. 10. Total Error in terms of number of iterations over a star network with 8 path branches of length 10.

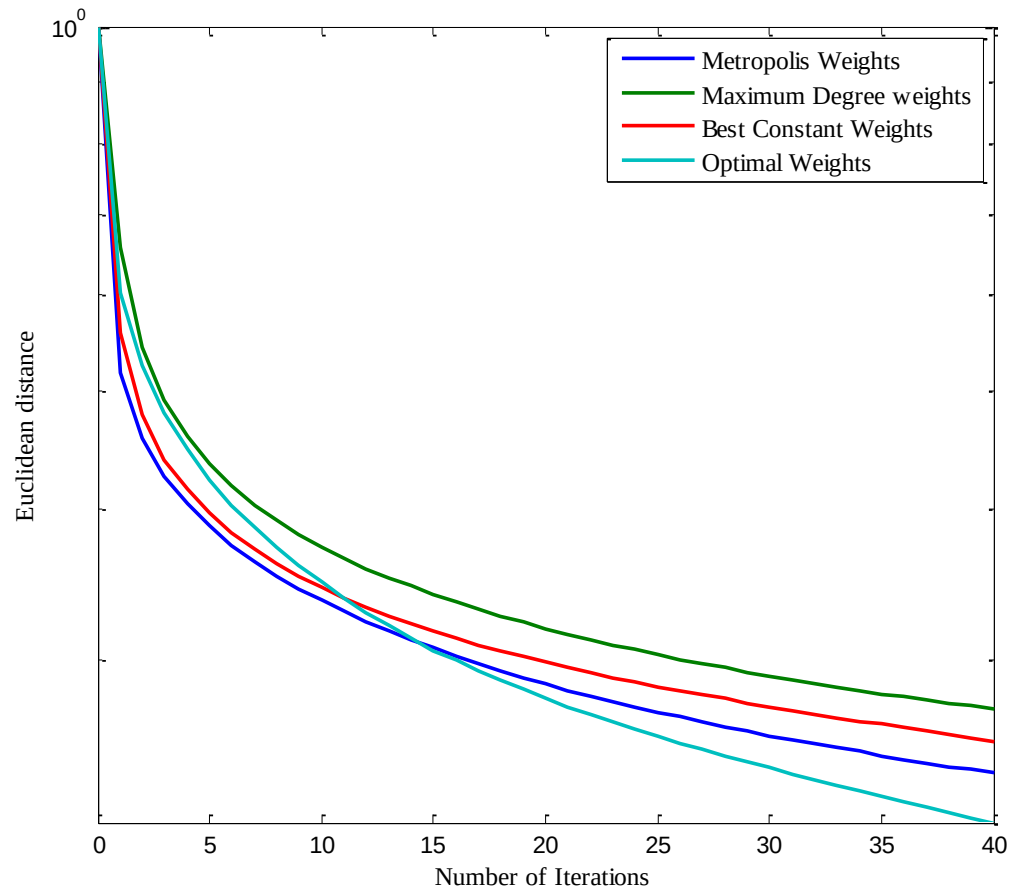

Fig. 11. Total Error in terms of number of iterations over a star network with 8 Lollipop branches of order (5,5).

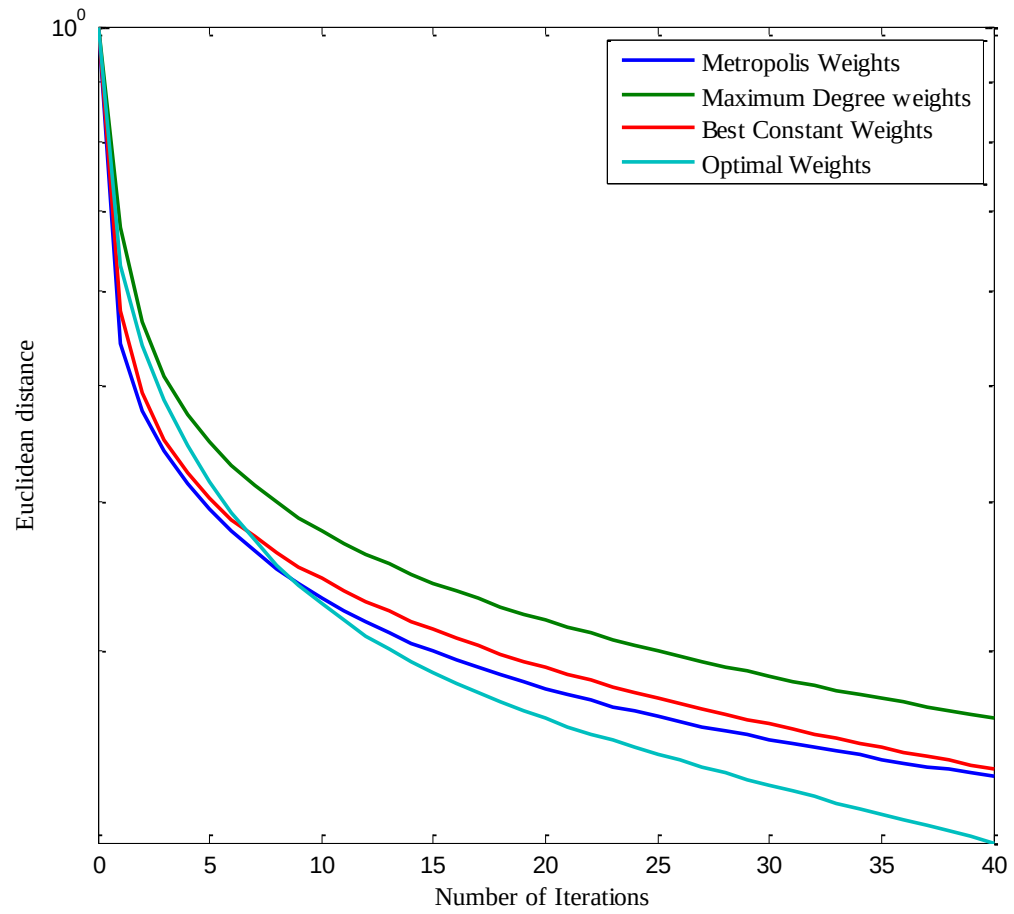

Fig. 12. Total Error in terms of number of iterations over a star network with 8 Semi-complete branches of order (5,2,3).

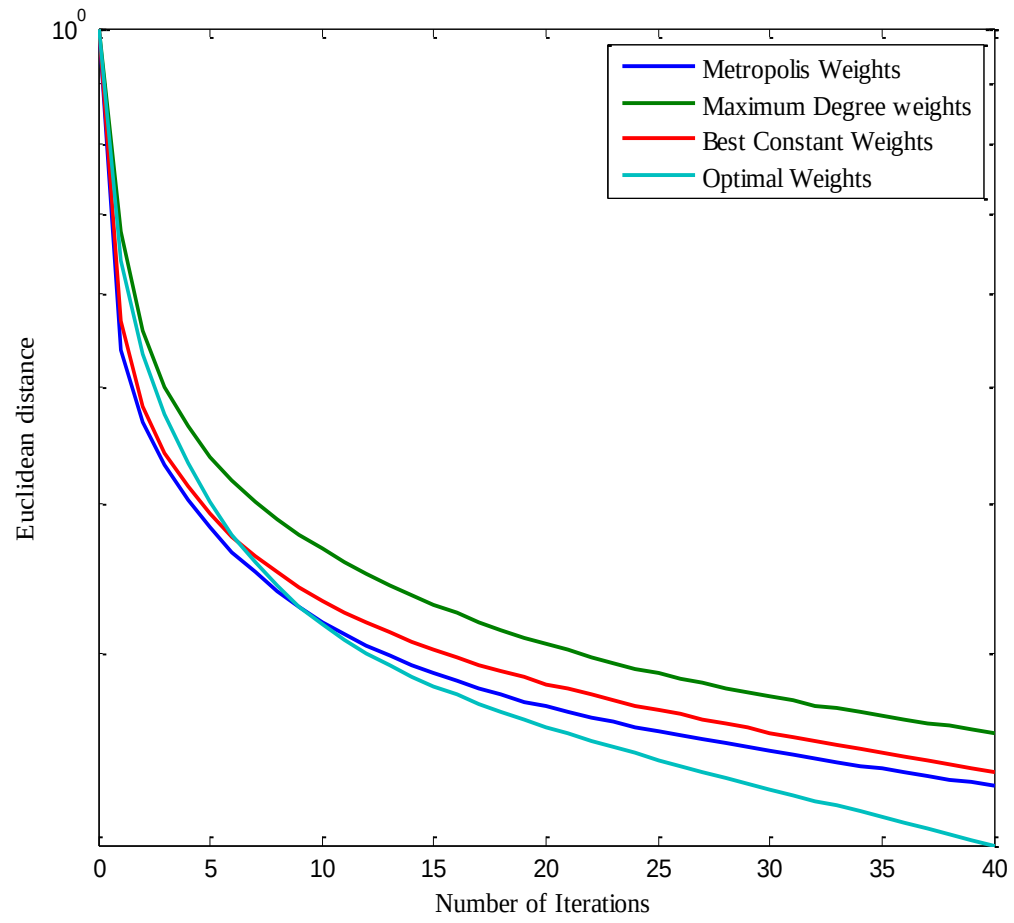

Fig. 13. Total Error in terms of number of iterations over a star network with 8 Ladder branches of order (1,2,2).

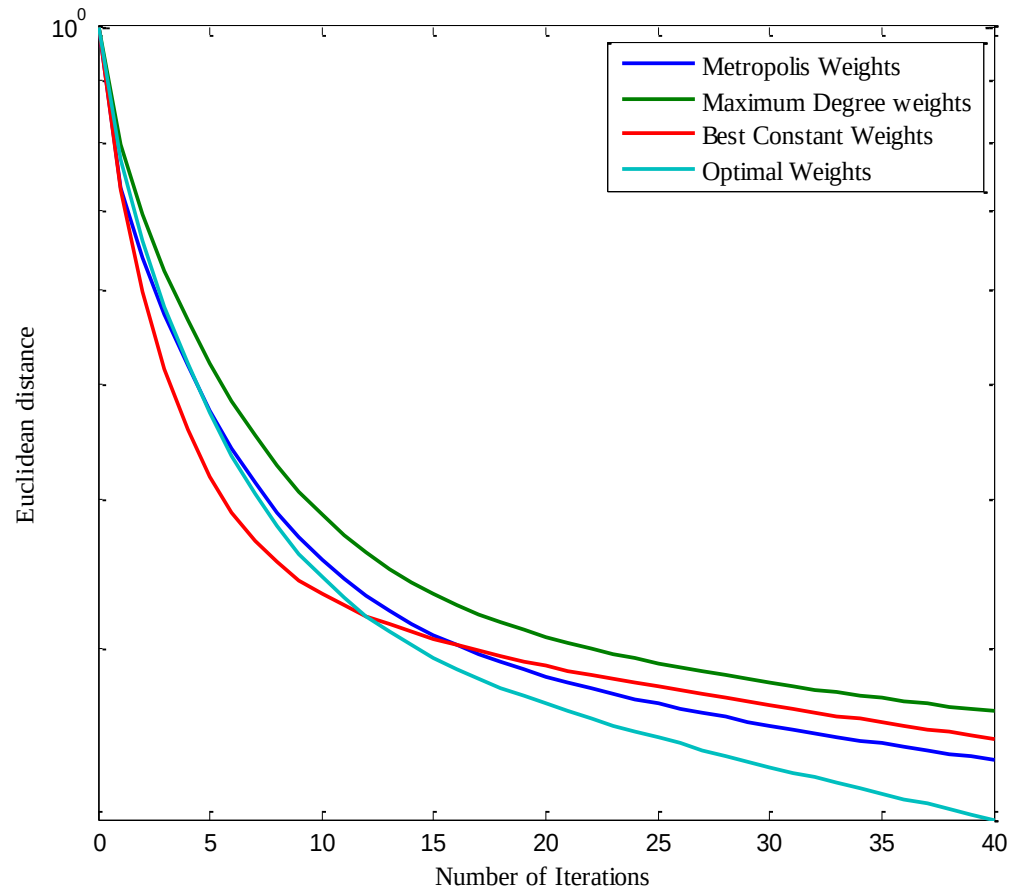

Fig. 14. Total Error in terms of number of iterations over a star network with 8 palm branches of order (4,5).

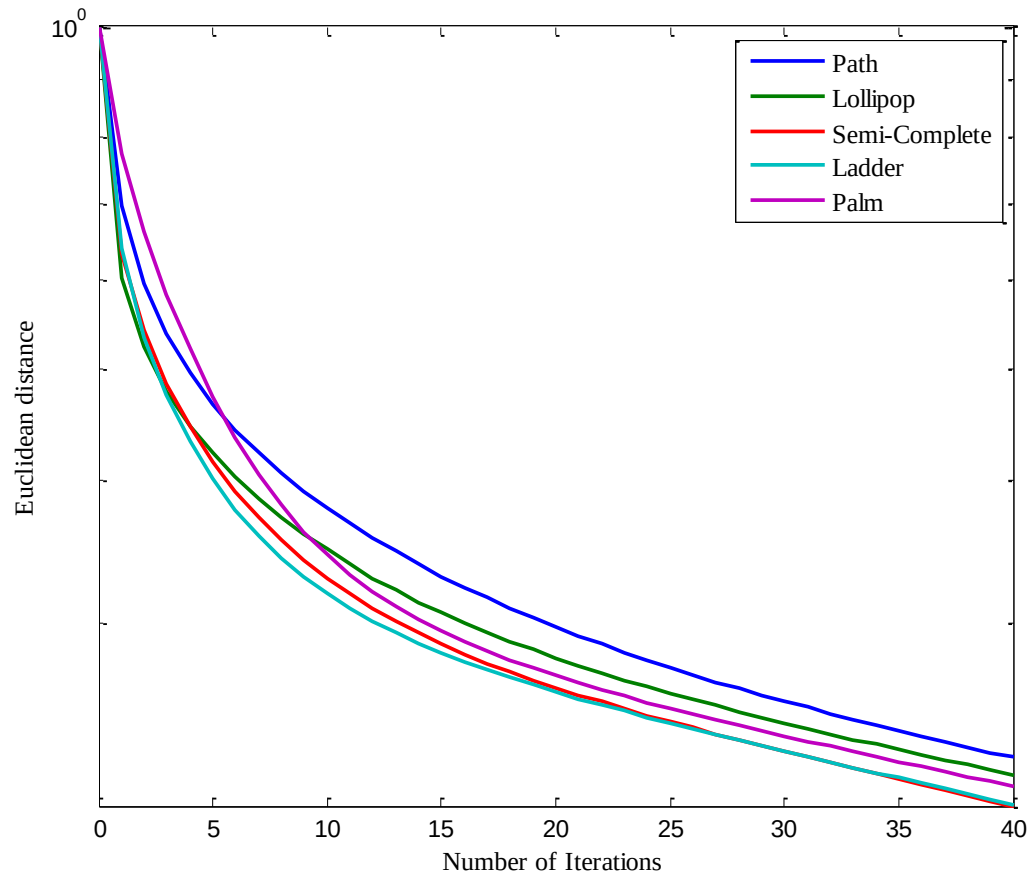

Fig. 15. Total Error in terms of number of iterations over a symmetric star network with 8 branches for the choice of Optimal Weights.

## VIII. Conclusion

Fastest Distributed Consensus averaging Algorithm in sensor networks has received renewed interest recently. In most of the methods proposed so far either numerical or analytical, full knowledge about the network's topology is required.

Here in this work, we have solved fastest distributed consensus averaging problem and determined the optimal weights for certain branches of an arbitrary connected network by means of stratification and semidefinite programming. We have shown that the obtained weights are independent of rest of the network and these weights can be used for branches of any connected sensor network. Our approach is based on fulfilling the slackness conditions, where the optimal weights are obtained by inductive comparing of the characteristic polynomials initiated by slackness conditions.

Examples and numerical results presented in paper confirm the optimality of obtained weights over other weighting methods. Moreover the obtained weights are optimal for combination of five branches introduced in paper. We believe that the method used for determining optimal weights is powerful and lucid enough to be extended to other types of branches with more general topologies, which is the object of future investigations. Other future directions include the addition of noise and considering the quantized data and communication delay in asynchronous mode.

## APPENDIX A
## MAXIMUM DEGREE, METROPOLIS-HASTING & BEST CONSTANT WEIGHTING METHODS

The Metropolis-Hastings weighting method is defined as:

$$W_{i,j} = \begin{cases} 1/\left(1 + max(d_i, d_j)\right) & j \in N_i, i \neq j \\ 1 - \sum_{j \in N_i} W_{i,j} & i = j \\ 0 & otherwise \end{cases}$$

where $d_i$ and $d_j$ are the degrees of nodes $i$ and $j$, respectively and $N_i$ is the set of immediate neighbors of node $i$.

The Maximum degree weighting method is defined as:

$$W_{i,j} = \begin{cases} 1/\max_k(d_k) & j \in N_i, i \neq j \\ 1 - d_i/\max_k(d_k) & i = j \\ 0 & otherwise \end{cases}$$

The best constant weighting method is defined as:

$$W_{i,j} = \begin{cases} \alpha & j \in N_i, i \neq j \\ 1 - d_i\alpha & i = j \\ 0 & otherwise \end{cases}$$

In [5] it has been shown that the optimum choice of $\alpha$ for best constant weighting method is $\alpha^* = 2/\left(\lambda_1(L) + \lambda_{n-1}(L)\right)$ where $\lambda_i(L)$ denotes the $i$-th largest eigenvalue of $L$ and $L$ is the Laplacian matrix defined as $L = AA^T$ with $A$ as the adjacency matrix of the sensor network's connectivity graph.